\documentclass[prb,twocolumn,showpacs,amsmath,superscriptaddress,amssymb]{revtex4}
\usepackage{graphicx}% Include figure files
\usepackage{dcolumn}% Align table columns on decimal point
\usepackage{bm}% bold math

\begin{document}

%\preprint{APS/123-QED}

\title{ Current control of systems with a Peierls distortion by magnetic field}% Force line breaks with \\
\author{Ryuji Takahashi}
\email{ryuji.takahashi@riken.jp}
%\affiliation{Department of Applied Physics, The University of Tokyo, Japan}
\affiliation{Condensed Matter Theory Laboratory, RIKEN, Wako, Saitama 351-0198,  Japan}
\author{Naoyuki Sugimoto}
\affiliation{Department of Applied Physics, The University of Tokyo, Japan}
\date{\today}% It is always \today, today,
    % but any date may be explicitly specified

\begin{abstract}
We study the tunneling phenomenon of a ladder system with a Peierls distortion in a magnetic flux,  and the response of electrons the insulator is investigated when the tunneling current flows on one-dimensional gapped chains along the external electric field.
Without the magnetic field, the ladder system is insulated by the charge density wave order.
Then, by the increase of the magnetic field, it becomes metallic with the disappearance of the distortion of the lattice.
 Finally, the gap appears, and it becomes a insulator.
 At the metallic state, the topological transition also occurs.
To show this phenomenon, we consider the distortion by the phonon in the ladder model, and calculate the distortion gap and the transition probability by using both Landau-Zener formula and the instanton method.
The transition to the metallic states will be applied to the current control by the magnetic field.
\end{abstract}

 % \pacs{71.30.+h, 72.10.Bg, 72.20.Ht, 73.22.Pr}% PACS, the Physics and Astronomy
        % Classification Scheme.
%\keywords{Suggested keywords}%Use showkeys class option if keyword
        %display desired
\maketitle
\section{Introduction}
Researches of one-dimensional (1D) quantum systems have been intensively done in condensed matter physics. 
Interactions of electrons can be simply treated in 1D systems with the various theoretical techniques \cite{Giamarchi}, and the effective 1D systems have been realized in experiments, e.g., carbon nanotube~\cite{Iijima91}, and organic molecules (TMTSF)$_2$PF\cite{Ishiguro,Jerome04}.
The Peierls transition refers the instability of the metallic states in 1D quantum wires\cite{Peierls}, and these are described by the Su-Schrieffer-Heeger (SSH) model\cite{Su79,Su80} in a simple way.
 In a half-filled band of the 1D chain, the dimerization occurs with the distortion of the lattice, and the gap is introduced at the Fermi energy.
%In addition, quantum ladders have been investigated intensively in the context of studies of the strongly correlated electrons \cite{Larkin93,Balents96}. Exchange of electrons can occurs without passing each other, and the magnetic field affects them.  Therefore, particles on the ladder with the mangetic field exhibits interesting phenomena \cite{Carr06}.

%Recently, the gapped band insulators have been studied in terms of topology in condensed matter physics\cite{Hasan10}.
%n addition to their topological states, transport phenomena have been intensively investigated \cite{Takahashi}.
These materials become insulating due to the gap, and the tunneling current is a one of the observation methods to study states of the matter.
 %To investigate the response of insulators by the external force, the quantum tunneling has been studied in a long time.
In a recent study, the enhancement of transition probability has been shown by taking into account the nonlinearity by the dispersion of the lattice~\cite{Takahashi17}. In a conventional way, the calculation of the transition probability is done by using the Landau-Zener (LZ) formula~\cite{Landau32,LandauB,Zener32}.
On the other hand, the nonlinearity peculiar to the periodic lattice affects the transition action, and this lattice effect becomes large when the transition probability is small.
Usually the probability is very small in insulating materials, and hence there is room for further research into the lattice effect on the transition phenomena in materials.
In addition, recent experiments have shown clear detection of the transition signal for the metal-insulator transition \cite{Asamitsu97, Miyano97}. 
In these systems, the transition of the LZ problem occurs on 1D chains parallel to the electric field.
 Then, the clear detection of the tunneling signal implies that the high density of the 1D chains in the material, and hence the hybridization between the 1D wires is expected.
By the hybridization,  the system is described by a two-leg ladder as a simple model, and the magnetic field affects on the transition dynamics of the electrons\cite{Carr06}.
 
In this paper, we consider spinless electrons in two coupled SSH chains with a magnetic field.
The Peierls states exhibit a gapped insulator, and the magnetic flux penetrates the system unlike the Meissner effect of the superconductors.
 Then, the level repulsion by the distortion is also expected by a magnetic field, similar to the Landau level.
However, we find that the distortion gap is reduced by the magnetic flux.
This means that the level attraction occurs by the magnetic field in the Peierls states.
In addition, the transition of the topological states occurs with the disappearance of the distortion.
In this case, the gap $\Delta$ disappears, and the velocity $v$ also vanishes at the transition point.

 At the same time, according to the LZ formula, the tunneling probability is given as $\sim\mathrm{e}^{-\pi\frac{\Delta^2}{vF}}$ with the external force $F$.
 Therefore, in the vicinity of the transition point, the behavior of the tunneling is elusive by the LZ formula, since the dispersion close to the Fermi energy is not linear.
The nonlinearity of the band structure becomes important, and the large lattice effect is expected \cite{Takahashi17,Takahashi18} due to $v=0$.
Then, we study the transition probability of the ladder system by using both of the LZ formula and the instanton method of the Bloch states \cite{Takahashi17}.
 Initially, when the magnetic field is absent, the ladder SSH system is insulating by the Peierls distortion.
 By increasing the magnetic flux, the system becomes metallic with the disappearance of the distortion; namely the transition probability becomes unity.
Then, the system becomes again insulating with the increase of the flux.
When the hybridization is large, it becomes strongly insulating in the large flux region,
 while it is weakly insulating for the small flux region since the distortion gap is small for the large hybridization.
We find that the calculation results by the two methods give similar behaviors in the vicinity of the transition point, since $\Delta$ rapidly vanishes  compared with $v$.
The probability by the instanton is larger than that of the LZ formula, and this is prominent for the large flux region with large values of $w$, since the lattice effect becomes strong with $w$.

 Our findings will be applied to current control by the magnetic field, since the rapid change of the transition probability occurs in the vicinity of the transition point. In addition, the Peierls distortion is related to the charged gap states\cite{McKenzie93}, and the insulator-metal transition can be obtained by tuning the magnitude of the hybridization between the two chains.
  One will obtain the metallic states from the insulator by various approaches.

\section{Ladder model in the magnetic field}
\begin{figure}[htbp]
 \begin{center}
 \includegraphics[width=80mm]{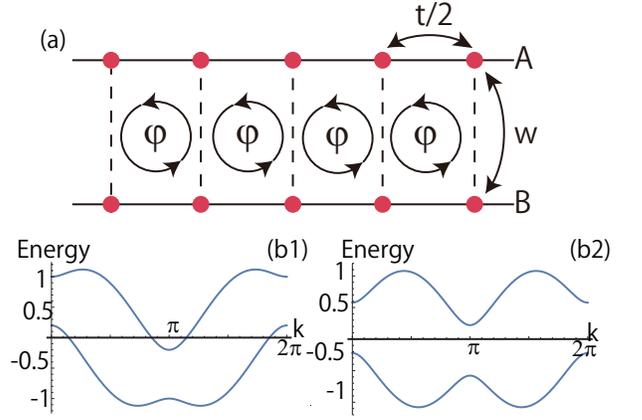}
 \caption{ (a)Schematic illustration of the ladder system threaded by the magnetic flux $\phi$, when the distortion is absent.
$\frac{1}{2}t$ is the hopping integral, and $w$ is the hybridization between the two chains.
(b1) (b2) Dispersion without the coupling between the electron and phonons for $\frac{w}{t}=0.4$, $\phi = 0.8\phi_{0}$ (b1), and $\phi = 1.2\phi_{0}$ (b2). The energy bands cross $E=0$ in (b1), while (b2) has the gap.
}
\label{fig:disp}
% \label{fig:RTfig}
\end{center}\end{figure}

We first consider a ladder system with a magnetic flux.
The Hamiltonian is given as
 \begin{eqnarray}
H_{0}&=&\frac{t}{2}\sum_{i,\mu=A,B}
(C^{\dagger}_{\mu, i +1}C_{\mu, i }\mathrm{e}^{-is_{\mu}\phi}+C^{\dagger}_{\mu, i-1}C_{\mu, i}\mathrm{e}^{is_{\mu}\phi})
\nonumber
\\
&\ &+w\sum_{i,\mu\not=\nu}C^{\dagger}_{\mu,i}C_{\nu,i }
\label{eq:h0}.
\end{eqnarray} 
where the first term is the Hamiltonian of the 1D chain labeled with $\mu=A,B$ (Fig.~\ref{fig:disp}(a)) with $s_{A(B)}= 1(-1)$, and $t$ is the hopping integral. The second term represents the interaction between the two chains with the strength $w$, and the half-filled band is considered, i.e., the Fermi energy $E_{F}=0$.
$\phi$ is the magnetic flux penetrating the ladder, and we consider $\phi\in[0,\frac{\pi}{2}]$ in this study, without loss of generality.
$C_{i}(C_{i}^{\dagger})$ is the annihilation (creation) operator at site $i$.
The non-interaction Hamiltonian (\ref{eq:h0}) is diagonalized by the unitary matrix
 \begin{eqnarray}
 U_{0}^{\dagger} H_{0}U_{0}&=&
 \begin{pmatrix}
E_{k}^{+} &0\\
0& E_{k}^{-}
\end{pmatrix},\\
 U_{0}&=&
 \begin{pmatrix}
\frac{w}{\sqrt{2\chi(\chi+t\sin k\sin\phi)}} & \frac{w}{\sqrt{2\chi(\chi-t\sin k\sin\phi)} }\\
\frac{\eta+t\sin k\sin\phi}{\sqrt{2\chi(\chi+t\sin k\sin\phi)}} & \frac{-\eta+t\sin k\sin\phi}{\sqrt{2\chi(\chi-t\sin k\sin\phi)}}
\end{pmatrix},\label{eq:u0}
 \end{eqnarray}
with $\chi=\sqrt{w^2+ t^2\sin^2 k\sin^2\phi}$.
The energy dispersion is given by $E^{\pm}_{k}$, with
 \begin{eqnarray}
E_{k}^{-}=E_{k}=t \cos k\cos\phi-\chi,
\label{eq:En}
\end{eqnarray}
and $E_{k}^{+}=-E_{k+\pi}$.
The degeneracy at $E=0$ appears for $\phi \leq \arccos\frac{w}{t}\equiv \phi_{0}$ as shown in Fig.~\ref{fig:disp}(b1), and the gap is expected by the interaction at the degeneracy.
On the other hand, for $\phi >\phi_{0}$, the system is gapped without the interaction (b2).

In the present paper, we study the gap brought by the interaction with the phonon at the degeneracy, i.e., the Peierls instability, when the degeneracy occurs. The interaction Hamiltonian has the form
 \begin{eqnarray}
H_{i}&=&\frac{1}{2}t\sum_{m,u,\mu}
(-1)^{u+1} \gamma_{\mu} C^{\dagger}_{\mu,2i+ u}C_{\mu,2i }\mathrm{e}^{-is_{\mu}\phi} +\mathrm{H.c.},
\end{eqnarray} 
where $u=\pm1$, and $ \gamma_{\mu}$ is the coupling strength between electron and phonon coming from the lattice distortion.
The distortion between the two chains can be assumed by the form $\sim cw\sum_{\mu\not=\nu ,i}
(-1)^{i}  C^{\dagger}_{\mu i}C_{\nu i }$ with the distortion strength $c$. 
However, in the present paper,
 we neglected this interaction, since it mainly affects the level repulsion between the higher (lower) energy bands.
 This can be seen by a unitary transformation of the Hamiltonian (\ref{eq:uniHam}).
% In Fig.~\ref{fig:disp}(c1), the dispersion is shown for $c=0.2$, $\gamma_{\mu}=0$, $\frac{w}{t}=0.4$, and $\phi=0$, and the level repulsion does not appear at $E=0$.
\subsection{Lattice distortion by the electron-phonon interaction}
\begin{figure}[htbp]
 \begin{center}
 \includegraphics[width=85mm]{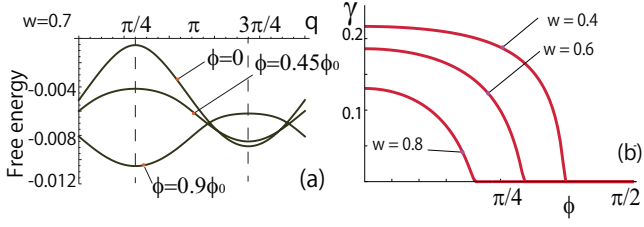}
 \caption{(a) The free energy in units of $t$ for $K=0$ as a function of the angle of the distortion $q$ for $\frac{\phi}{\phi_{0}}=0,0.45,0.9$, at $w=0.7$.
 The angle is given by $ (\gamma_{A}, \gamma_{B})=r(\cos q,\sin q)$, and we use $r=0$.
 For $q=\frac{3\pi}{4}$ we have the saddle point of the free energy i.e., $F_{e} (\gamma, -\gamma)<F_{e} (\gamma, \gamma)$ for the small flux.
When the flux becomes large, we have the saddle point for the configuration $(\gamma, \gamma)$.
(b)The distortion $\gamma$ as a function of the flux $\phi$.
 $ \gamma $ decreases with $\phi$ and vanishes when $\cos\phi=\frac{w}{t}$ .
}
\label{fig:alph}
% \label{fig:RTfig}
\end{center}\end{figure}

By the Hamiltonian $H=H_{0}+H_{i}$, we have the four eigenvalues $\pm\xi^{l}_{k}$ and $\pm\xi^{ll}_{k}$, with $\xi^{l,ll}>0$, 
and the free energy per unit volume is expressed as
 \begin{eqnarray}
 F_{e} (\gamma_{A}, \gamma_{B})=-\frac{1}{N}\sum_{k}\xi_{k}^{t} +\frac{K}{2}(\gamma_{A}^2+\gamma_{B}^2),\label{eq:fen}
\end{eqnarray} 
where $K$ is the elastic constant, and $\xi_{k}^{t}=\xi^{l}_{k}+\xi^{ll}_{k}$, and $N$ is the number of the unit cells.
By the symmetry of the Hamiltonian, we have relations $F_{e} (\gamma_{A}, \gamma_{B})=F_{e} (-\gamma_{A}, -\gamma_{B})$, and $F_{e} (\gamma_{A}, \gamma_{B})=F_{e} (\gamma_{B}, \gamma_{A})$.
 Therefore, $\gamma_{B} = \gamma_{A}$ and $\gamma_{B} = -\gamma_{A}$ are candidates for the saddle point.
By the calculation, we obtain the minimum energy for $\gamma_{B} = -\gamma_{A}$ when $\phi$ is small. For the large $\phi$, the minimum energy appears at $\gamma_{B} = \gamma_{A}$.
In Fig.~\ref{fig:alph}(a), the free energy with $K=0$ is shown for $\frac{\phi}{\phi_{0}}=0,0.45,0.9$ at $w=0.7$ by representing $ (\gamma_{A}, \gamma_{B})=r(\cos q,\sin q)$.
%Figure \ref{fig:disp}(c2) shows the energy dispersion for $c=0$, $\gamma_{A}=0.2$, $\gamma_{B}=0.2$, $\frac{w}{t}=0.4$, and $\phi=0$; the system is metallic.
 For the small flux, the result is consistent with the earlier study of the 2D Peierls instability with $\phi=0$~\cite{Hirsch88}.
Therefore, the change of the saddle point occurs by increase of the flux.
This change of the saddle point by the magnetic field is seen by the effective gap.
By assuming $ \gamma \ll1$, the gap $\Delta(\gamma_{A}, \gamma_{B})$ by the instability is given as (Eq.~\ref{eq:efgap})
 \begin{eqnarray}
\Delta(\gamma, \gamma)&=&\frac{\gamma t\sin\phi}{\sqrt{w^2+t^2\sin^2\phi}},\\ 
\Delta(\gamma, -\gamma) &=&\frac{\gamma \sqrt{t^2\cos^2\phi-w^2}}{\cos \phi}\label{eq:effgap}.
\end{eqnarray} 
We have $\Delta(\gamma, -\gamma)>\Delta(\gamma, \gamma)$ for small values of $\phi$, and the energy gain is maximized for $\gamma_{B}= -\gamma_{A}$.
 On the other hand, we have $\Delta(\gamma, -\gamma)<\Delta(\gamma, \gamma)$ when the magnetic field is large, i.e., $\sqrt{\cos2\phi}< w(<t\cos\phi)$.
 Then, there is a problem if the configuration $\gamma_{A}=\gamma_{B}$ is actually realized.

 To answer the question, we consider the e-ph interaction,
  \begin{eqnarray}
  H_{\mathrm{e-ph}}=\sum_{q, k} C^{\dagger}_{ k+q }G_{q}C_{ k }+\mathrm{H.c.},
 \end{eqnarray}
where $C_{ k }=(C_{A, k },C_{B, k })$ and $G_{q}=\mathrm{diag}(G_{A,q},G_{B,q})$ is the magnitude of the interaction.
In this case, the microscopic Hamiltonian has the form,
$H=H_{0}+H_{\mathrm{e-ph}}$.
As shown previously, $H_{0}$ is diagonalized by $U_{0}$ (Eq.~(\ref{eq:u0})), 
and this transformation affects the interaction $H_{\mathrm{e-ph}}$.
 By using $U_{0}$, we express the electron operators as $\Tilde{C}_{k } =U_{0, k}^{\dagger}C_{k }$.
 Then, we have
  \begin{eqnarray}
  H_{\mathrm{e-ph}}=\sum_{q, k}\Tilde{C}^{\dagger}_{ k+q } U_{0, k+q}^{\dagger}G_{q} U_{0, k} \Tilde{C}_{k }+\mathrm{H.c.}.
 \end{eqnarray}
 Here, we extract the relevant interaction, i.e., $q=\pi$,
  \begin{eqnarray}
  H_{\mathrm{e-ph}}\sim \Tilde{C}^{\dagger}_{k+\pi } U_{0, k+\pi}^{\dagger}G_{\pi} U_{0, k} \Tilde{C}_{ k }+\mathrm{H.c.} \nonumber.
 \end{eqnarray}
The offdiagonal elements (interband interaction) give the relevant interaction according to the previous analysis, and it has the form
  \begin{eqnarray}
  H_{\mathrm{e-ph}}\sim (G_{A,\pi}-G_{B,\pi})\sigma_{x},
 \end{eqnarray}
 with the Pauli matrix $\sigma_{x}$.
Therefore, the magnitude of the relevant e-ph interaction is given by the difference of the respective interactions.
Here, the magnitude of the interaction is proportional to the distortion, i.e., $G_{\mu,\pi}\propto \gamma_{\mu}$, since $H_{i}$ originates from the e-ph interaction.
By taking into account the free energy analysis, we conclude that the configuration $\gamma_{A}= \gamma_{B}$ does not appear by the e-ph interaction, although the configuration $\gamma_{A}= \gamma_{B}$ gives a stable state of the free energy when the magnetic field is strong.
%This may be obtained for the electron-electron interaction.
Therefore, we use $\gamma_{A}=-\gamma_{B}= \gamma $ in the following discussion.

\begin{figure}[htbp]
 \begin{center}
 \includegraphics[width=60mm]{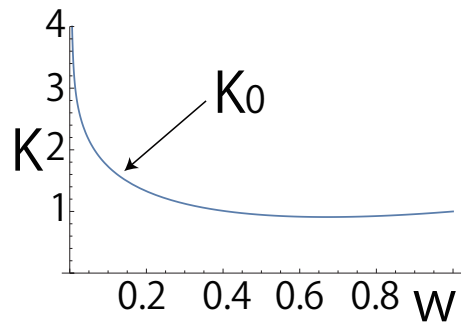}
 \caption{
Plot of $K_{0}$ as a function of the hybridization $w$ in units of $t$.
For $K>K_{0}$, $\partial_{ \gamma }^2F>0$ is satisfied at $w=t\cos\phi$.
}
\label{fig:ksl}
% \label{fig:RTfig}
\end{center}\end{figure}

By maximizing the energy gain with respect to $ \gamma $, e.g., $\partial_{ \gamma }F_{e}=0$ and $\partial_{ \gamma}^2F_{e}>0$,
we obtain the stable configuration.
In addition, when the degeneracy at $E=0$ is lost, we assume that the coupling vanishes, since the gap is mediated by the scattering in the degeneracy.
In other words, when the gap is absent, the distortion should vanish; namely we have to have $\gamma =0$ at $t\cos\phi =w $.
Then, a restriction on $K$ appears by the curvature of the free energy for $\gamma=0$ at $\phi=\phi_{0}$.
To show this, we consider the free energy for $\gamma\ll1$ by using the effective Hamiltonian in $k$-space (appendix~\ref{app:effH}),
 \begin{eqnarray}
h_{k}^{\pm}&=&\begin{pmatrix}
\pm E_{k}^{\pm} &-  \gamma  i\partial_{k}E_{k}^{\pm}\\
i \gamma \partial_{k}E_{k}^{\pm} &\mp E_{k}^{\pm}
\end{pmatrix}.\label{eq:efHam}
\end{eqnarray}
Since the two Hamiltonians are symmetric, $h_{k}^{-} =-h_{k+\pi}^{+}$, we consider one of them $h_{k} \equiv h_{k}^{-}$.
The effective free energy per one atom is given as
 \begin{eqnarray}
 f_{e} =-\frac{1}{N}\sum_{k}\xi_{k} +\frac{K}{2}\gamma^2,
\end{eqnarray} 
with $\xi=\sqrt{E_{k}^2 + \gamma ^2(\partial_{k}E_{k})^2 }$.
By $\partial_{ \gamma }^2 f|_{ \gamma \to0} $, the condition reduces to
 \begin{eqnarray}
K>K_{0}\equiv \sum_{k}\frac{(\partial_{k}E_{k})^2}{| E_{k}|}.
\end{eqnarray} 
The calculation result of $K_{0}$ is shown in Fig.~\ref{fig:ksl}, and we use $K>K_{0}$ in the following discussion, e.g., $K \sim 24t$ in polyacetylene \cite{polyacetylene}.
We note that the electronic states in the vicinity of the Fermi energy cannot be described by the Tomonaga-Luttinger model at the transition point, since the curvature effects become important.

Under the restriction, we show the distortion $ \gamma $ with respect to the magnetic flux $\phi$ by the free energy(Fig.~\ref{fig:alph})(b).
The results are shown for $\frac{w}{t}=0.4,0.6, 0.8$ at $K=1.2t$.
By increasing $\phi$, $ \gamma $ becomes small, and vanishes at $\phi_{0}=\arccos\frac{w}{t}$ with $0<\phi<\frac{\pi}{2}$.
For small $w$, $\gamma$ decreases slowly for small values of $\phi$, and rapidly changes in the vicinity of $\phi_{0}$.
Due to $ \gamma ^2\ll1$, the band edge $k_{0}$ is approximately given by the condition $E_{k_{0}}=0$, i.e., $\cos k_{0}=\frac{\sqrt{w^2+t^2\sin^2\phi}}{t}$.
 The effective gap is given by Eq.~(\ref{eq:effgap}).

\subsection{Topological states of the ladder model}
Here, we briefly mention topological states of the ladder model.
The SSH model is a 1D topological material which has the BDI symmetry\cite{Schnyder08}.
The ladder Hamiltonian in k-space is written as
 \begin{eqnarray}
H_{k}&=&t\cos k\cos\phi\sigma_{x}-t\sin k\sin\phi\sigma_{x}\tau_{z}\nonumber\\
&\ &-\gamma t\sin k\cos\phi\sigma_{y}-\gamma t\cos k\sin\phi \sigma_{y}\tau_{z}+w\tau_{x},\nonumber\\
\end{eqnarray}
where $\boldsymbol{\sigma}$ and $\boldsymbol{\tau}$ are the Pauli matrices for the sublattice and ladder degrees of freedom.
The above Hamiltonian has the chiral symmetry, $H_{k}=-\sigma_{z}\tau_{z}H_{k}\sigma_{z}\tau_{z}$.
In addition, it also has a symmetry by C$_{2}$ rotational transformation: $H_{k}=\tau_{x}H_{-k}^{t}\tau_{x}$.
Therefore, the ladder system belongs to the BDI symmetry class, and has a nontrivial topological number, according to Ref.~\onlinecite{Song17}.
In the present model, when the distortion disappears, the Hamiltonian is given as
 \begin{eqnarray}
H_{k}&=&t\cos k\cos\phi-t\sin k\sin \phi \tau_{z} + w\tau_{x}.
\end{eqnarray}
Then, the system becomes trivial\cite{Ryu02,Takahashi13}, and therefore, the topological transition occurs at $t\cos\phi =w$.

\section{Transition probability}
Having established the ladder model in the magnetic field, we consider the transition probability using the whole Hamiltonian 
 \begin{eqnarray}
H_{0}+H_{i}+V(x),
\end{eqnarray}
where the potential is proportional to the position operator, $V(x)=\sum_{i,\mu=A,B } x FC_{\mu i}^{\dagger }C_{\mu i}$ with the force $F$.
To this end, we calculate the transition probability based on the instanton method of the Bloch states\cite{Takahashi17}.
 In this model, by increasing the magnetic flux $\phi$, the gap by the distortion vanishes at $\phi_{0}=\arccos\frac{w}{t}$, and metallic states appear. Then, the system becomes gapped again by increasing $\phi$; the hybridization $w$ plays a roll of the gap instead of the distortion.
Therefore, we separately consider the system with $\phi\leq \phi_{0}=\arccos\frac{w}{t}$ and $\phi>\phi_{0}$. 
\subsection{Action for small magnetic field}
%According to the study on the tunneling problem~\cite{Takahashi17}, we calculate the transition probability of the ladder system.
For small magnetic field $\phi\leq \phi_{0}$, the lattice distortion is present, and plays a role of the gap.
To obtain the transition path of the instanton, we consider the wavenuber on the 2D complex plane $k\to z=x+iy$.
 The condition for the branch cut $C$ is given by $E_{z=z_{c}}^2\leq 0$, which reduces to
 \begin{eqnarray}
\cos^2 [x_{i}(y)] %&=&\sin^2\phi+\frac{w^2}{v^2} \frac{1}{\cosh^2y +\sinh^2 y\tan^2\phi}\nonumber\\
&=&\sin^2\phi+\frac{w^2}{v^2} \frac{\cos^2\phi}{\sinh^2 y+\cos^2\phi},
\label{eq:bc}
\end{eqnarray}
 where we have the band edge $x_{i}(0)=k_{0}$ for $y=0$.
Since the gap is small, we use the effective gap (\ref{eq:effgap}), and 
%\sin^2 [x_{i}(y)] \sinh^2 y= \frac{\Delta^2}{v^2}\cos^2 \phi\label{eq:bp}
the equation $E_{z}^2+\Delta_{0}^2=0$ gives the branch points $y_{0}^{\pm}$ as
\begin{eqnarray}%
y_{0}^{\pm}&=&\pm y_{0}= \pm \mathrm{arcsinh}
\sqrt{R^2+\sqrt{ R^4+ \frac{\Delta_{0}^2\cos^2\phi}{t^2 } }}
\end{eqnarray} 
with
$R=\sqrt{\frac{w^2+\Delta_{0}^2 -t^2\cos^2\phi}{2t^2} }$.
Then, the instanton action is given as
 \begin{eqnarray}
iS&=&-\frac{1}{F}\int_{C} \mathrm{d }k ~\xi_{z}=-\frac{2}{F}\int^{y_{0}}_{0} \mathrm{d }k ~\xi_{y} \label{eq:acsm}.
\end{eqnarray}
In addition, by the $k\cdot p$ approximation (the LZ formula), the classical action is given as
\begin{eqnarray}
iS_{k\cdot p}=-\frac{\pi}{2}\frac{ t}{F} \gamma^2\label{eq:apacsm}.
\end{eqnarray}
Since the effective gap (\ref{eq:effgap}) is proportional to the velocity, the LZ formula of the ladder model does not have a singularity at the transition point.

\subsection{Large magnetic field}
For large magnetic field $\phi>\phi_{0}$, the gap by the lattice distortion is absent.
The energy dispersion is described by the Hamiltonian $H_{0}=\mathrm{diag}(E_{k}^{+}, E_{k}^{-})$,
 and the transition occurs between the two states.
 In this case, the band edge appears at $k_{0}=0,\pi$ in the Brillouin zone.
 However, the energy difference at these points $E_{k_{0}}^{+}-E_{k_{0}}^{-}=2w$ is not the smallest gap;
 the energy difference between the band edges on $k=0$ and $k=\pi$ is the smallest, i.e., $E_{k=\pi}^{+} -E_{k=0}^{-}=2(w-t\cos\phi)$.
%the transition between the smallest energy difference ($E_{k=\pi}^{+}$ and $E_{k=0}^{-}$) does not belong to the Landau-Zener problem.
We first consider the energy transition between $E_{\pi}^{+}$ and $E_{0}^{-}$ by the LZ problem of the lattice.
To describe this transition, we assume a scattering between the two states with the energies $E_{k=\pi}^{+}$ and $E_{k=0}^{-}$: $V_{k}=\eta \phi_{+,k+\pi}^{\dagger}\phi_{-,k} +\mathrm{H.c.}$ with the eigenstate $\phi_{\pm,k}$ of $H_{0}$.
 This gives the Hamiltonian $\Tilde{H}_{k}=\mathrm{diag}(\Tilde{\xi}_{k} -\Tilde{\xi}_{k})$, with $\Tilde{\xi}_{k}=\sqrt{ E_{k}^2+\eta^2}$.
 By taking the limit $\eta\to0$, we find that the instanton path is given by $x=0$.
 Then, we obtain the instanton action
 \begin{eqnarray}
iS &=&-\lim_{\eta\to0}\frac{1}{F}\int_{C} \mathrm{d }k ~\Tilde{\xi}_{k}
=\frac{2}{F}\int_{0}^{y_{0}} \mathrm{d }y E_{iy}\nonumber\\
&=&\frac{2t\cos\phi}{F}\sqrt{\left(\frac{w}{t}\right)^2-\cos^2\phi} \nonumber\\
&\ &- \frac{2w\sqrt{\rho^2+1}}{F}\int_{\lambda_{0}}^{\frac{\pi}{2}}\mathrm{d}\lambda \partial_{\lambda} G\left(\lambda, \sqrt{\frac{1}{\rho^2+1}}\right),\label{eq:aclm}
\end{eqnarray}
 \begin{eqnarray}
 G\left(\lambda, \sqrt{\frac{1}{\rho^2+1}}\right) &=&F\left(\lambda, \sqrt{\frac{1}{\rho^2+1}}\right)+E\left(\lambda, \sqrt{\frac{1}{\rho^2+1}}\right),
\nonumber\\
\end{eqnarray}
where the functions $F(a,b)$ and $E(a,b)$ are the incomplete elliptic integral of the first and second kinds, and
$y_{0}=\mathrm{arcsinh} \sqrt{\left(\frac{w}{t}\right)^2-\cos^2\phi} $ (see appendix~\ref{app:cal}).
By the $k\cdot p$ approximation, we have $E_{k}\sim t\cos\phi-w-t\left(\cos\phi+\frac{t}{w}\sin^2\phi\right)\frac{k^2}{2}$ (Eq.~(\ref{eq:En})), and the transition action is given as 
  \begin{eqnarray}
iS_{k\cdot p}&=&-\frac{4}{3F}\sqrt{\frac{2(w-t\cos\phi )^3}{ t\left( \cos\phi+\frac{t}{w}\sin^2\phi \right)}} \label{eq:apaclm}.
\end{eqnarray}
The above action is different from that of the LZ formula.
We note that the tunneling probability by the energy difference $E_{0(\pi)}^{+}-E_{0(\pi)}^{-}$ is smaller than the probability by the above results for $\phi<\frac{\pi}{2}$ (see appendix~\ref{ap:transpi}).

\subsection{Calculation results of the tunneling current}
\begin{figure}[htbp]
 \begin{center}
 \includegraphics[width=80mm]{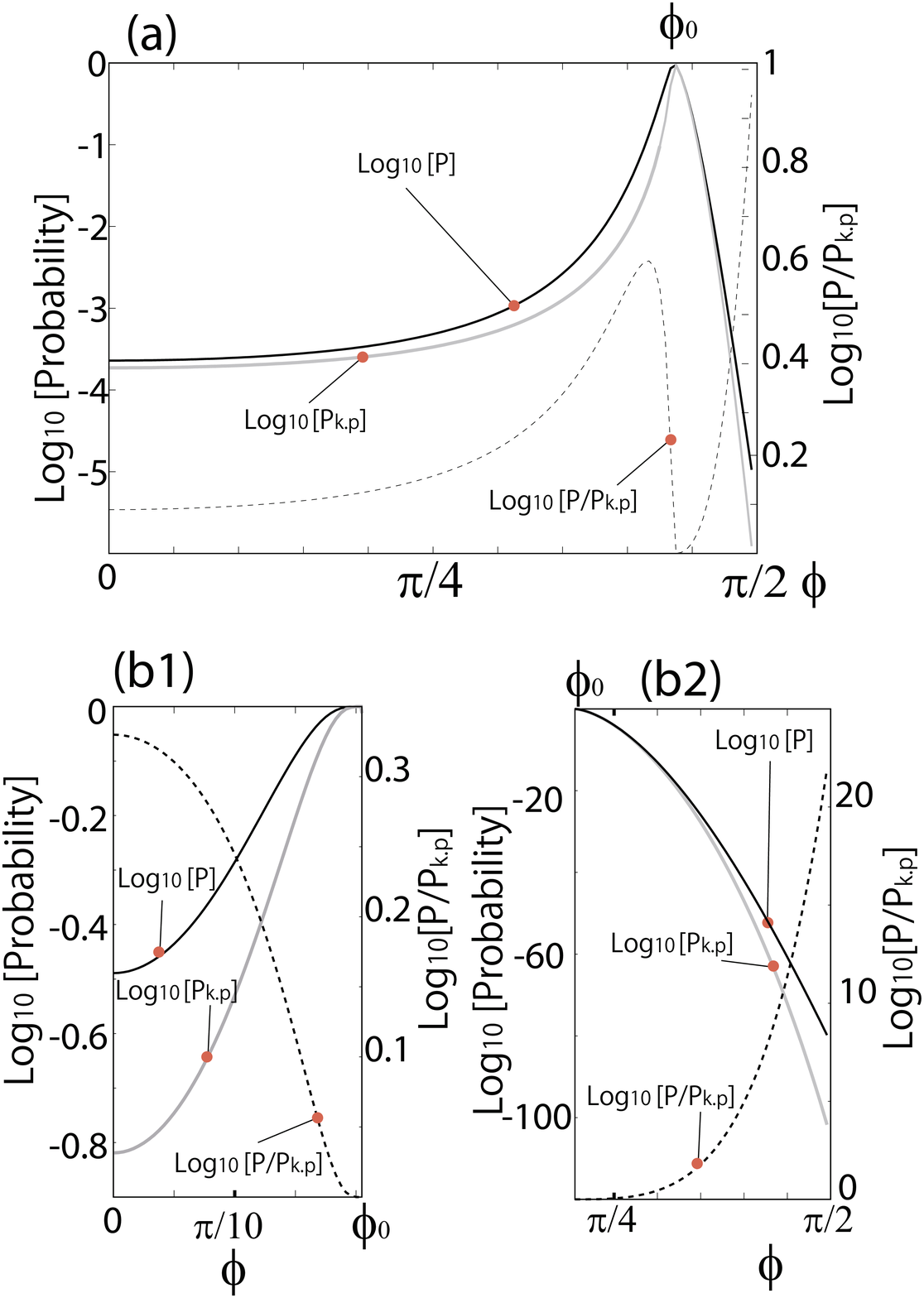}
 \caption{Transition probabilities by the instanton calculation $P$ and approximation methods $P_{\mathrm{k\cdot p}}$ in a log scale at $F=0.01$, for $w=0.2$(a) and $w=0.8$(b1)(b2).
The solid curve shows the probability by the instanton method, and the gray curve shows the probability by the approximation.
Their magnitude is shown on the left vertical axis.
 The dotted curve is express the ratio of the two probability, and the magnitude is shown on the left vertical axis.
} 
\label{fig:prob}
% \label{fig:RTfig}
\end{center}\end{figure}

By using the instanton action in the previous section, we calculate the transition probability
  \begin{eqnarray}
P=\mathrm{e}^{iS}.
\end{eqnarray}
The calculation results of the tunneling probability are shown in Figs.~\ref{fig:prob}(a),(b1),(b2),
 for the small and large hybridization $w=0.2$(a) and $w=0.8$(b1)(b2).
 The magnitude of the probability is shown in the left vertical axis.
 The solid curve shows the probability by the instanton action (Eqs.~(\ref{eq:acsm})(\ref{eq:aclm})).
The probability by the approximated action (Eqs.~(\ref{eq:apacsm})(\ref{eq:apaclm})) is expressed as $
P_{\mathrm{k\cdot p}}=\mathrm{e}^{iS_{\mathrm{k\cdot p}}}$, and is shown by the gray curve.
The ratio $\frac{P}{P_{\mathrm{k\cdot p}}}$ is shown by the dotted curve with the magnitude on the right vertical axis.
As the action shows, the probability $P$ increases with $\phi$ in the small $\phi$ region, and at the transition point $\phi=\phi_{0}$ we have $P=1$ due to the disappearance of the distortion $\gamma$ and $w-t\cos\phi=0$.
 Then, in the large $\phi$ region, 
$P$ becomes small with the increase of $\phi$, due to the increase of the energy difference $\propto w-t\cos\phi$.

For the small $w$ (Fig~\ref{fig:prob}(a)), the calculation result of the instanton is well described by the approximated action. 
By varying the flux, initially the change of the transition is moderate, since the dependence of $\gamma $ on $\phi$ is small for small values of $w$.
The difference between the two calculation results by the instanton and $k.p$ methods originates from the nonlinearity of the energy dispersion by the periodic nature of the lattice.
This lattice effect is enhanced by increasing $w$ or $ \phi$, since the velocity is $ \propto \frac{\sqrt{t^2\cos^2\phi-w^2}}{\cos\phi} $. Then, the difference is small when the two parameters are small, and increase by varying $\phi$.
% by the group velocity
%On the other hand, the gap by the distortion decreases by the increase of $\phi$, and the probability becomes unity on the transition point $\phi_{0}$ for both calculation results. Therefore, the maximum value of the ratio appears as shown in (a),
For the large hybridization, in the small $\phi$ region, the difference is also small (b1), since the lattice effect emerges when the probability becomes small \cite{Takahashi17}.
For the large $\phi$ (b2), we obtain the large difference between the two results as shown in (b2).
These reflect the enhancement of the nonlinearity by the large values of $w$ and $\phi$, and the probability becomes small by it.
% This effect becomes prominent since the energy difference becomes large.
%As the study of the 1D chain \cite{Takahashi17}, the lattice effect emerges when the probability becomes small.
Consequently, in this study, the transition probability is well described by the approximated results in a qualitative manner.

\section{Summary and Discussion}
In our study, we consider the tunneling phenomenon of a ladder system with the Peierls distortion in a magnetic field.
To describe the system, we consider two 1D chains with the interaction, and we calculate the free energy to obtain the distortion gap.
Then, the transition probability is calculated according to the instanton method in the lattice \cite{Takahashi17}.
In the small flux region, the system has topologically nontrivial edge states. In reality, the material with the high density ladder system probably contains the topological mid gap states. 
Then, the transition via the mid gap states occur \cite{Sugimoto08}, and the transition probability will be enhanced for this region compared with our calculation results.
%In our theory, we did not consider the transition phase and adiabatic phase, to focus on the transition phenomena by the magnetic field.

The transition point $\phi_{0}=\arccos\frac{w}{t}$ is simply determined by parameters of the band structure, and it does not depend on the spring constant.
In addition, $\phi_{0}$ will be controlled by the pressure \cite{RibeiroRM09}; $w$ is tuned by varying the length between the two wires.
Then, we can control the current by a slight change of the magnetic field since the tunneling probability is sensitive in the vicinity of $\phi$.
 Therefore, this phenomenon will be expected to various application in semiconductors physics.

%\begin{acknowledgments}
%RT was partially supported by Grant-in-Aid for Japan
%Society for the Promotion of Science Fellows (No.~25-9798).
%\end{acknowledgments}

\appendix
\section{Effective Hamiltonian of the ladder model}\label{app:effH}
The Hamiltonian of the ladder model with the distortion $(a,b,c)$ $H_{0}+H_{i}$ is represented as
 \begin{eqnarray}
H_{k}
&=&
\begin{pmatrix}
h_{1}&V^{\dagger}\\
V&h_{2}
\end{pmatrix},
\end{eqnarray}
where
\begin{eqnarray}
h_{1}&=&
\begin{pmatrix}
t\cos k\cos\phi+w&\sin k \sin\phi\\
t\sin k \sin\phi& t\cos k\cos\phi-w \\
\end{pmatrix},\\
h_{2}&=&
\begin{pmatrix}
- t\cos k\cos\phi+w& -t\sin k \sin\phi\\
 -t\sin k \sin\phi&- t\cos k\cos\phi-w \\
\end{pmatrix}
\end{eqnarray}
and we used the unitary transformation $U_{1}$ for $H_{0}+H_{i}$ in k-space with
 \begin{eqnarray}
U_{1}
&=&\frac{1}{2}
\begin{pmatrix}
1&1&1&1\\
-1&-1&1&1\\
-1&1&-1&1\\
1&-1&-1&1
\end{pmatrix}.
\end{eqnarray}
$V$ is given as
 \begin{eqnarray}
V&=&
\begin{pmatrix}
\frac{i}{2}(\gamma_{A} f_{y}+\gamma_{B} f_{y})- cw & -\frac{i}{2}(\gamma_{A} f_{y}-\gamma_{B} f_{y})\\
-\frac{i}{2}(\gamma_{A} f_{y}-\gamma_{B} f_{y})&\frac{i}{2}(\gamma_{A} f_{y}+\gamma_{B} f_{y})+ cw 
\end{pmatrix}\nonumber\\
&=&\frac{i}{2}(\gamma_{A} f_{y}+\gamma_{B} f_{y}) - cw \sigma_{z}-\frac{i}{2}(\gamma_{A} f_{y}-\gamma_{B} f_{y})\sigma_{x}.
\end{eqnarray}
where $f_{y}=t\sin(\phi+k)$, and $g_{y}=t\sin(\phi-k)$.
Then, $V$ is treated as a perturbation for small distortion.
By using the unitary transformation $U$
 \begin{eqnarray}
UHU^{\dagger}=
\begin{pmatrix}
u^{\dagger}h_{1}u&u^{\dagger}V^{\dagger}u^{\dagger}\\
u Vu &uh_{2}u\label{eq:uniHam}
\end{pmatrix},
\end{eqnarray}
where
\begin{eqnarray}
U=\begin{pmatrix} 
u^{\dagger}&0\\
0&u
\end{pmatrix}
\end{eqnarray}
with $u=\cos\frac{\theta}{2}-i\sin\frac{\theta}{2}\sigma_{y}$.
For $\cos\theta=\frac{w}{\sqrt{w^2+ t^2\sin^2 k\sin^2\phi}}$ and $\sin\theta=\frac{t\sin k\sin\phi}{\sqrt{w^2+ t^2\sin^2 k\sin^2\phi}}$, we obtain 
 \begin{eqnarray}
&\ &u^{\dagger}h_{1}u\nonumber\\&=&t\cos k\cos\phi \nonumber\\ &\ &+
\begin{pmatrix}
\sqrt{w^2+ t^2\sin^2 k\sin^2\phi} &
0
\\
0&-\sqrt{w^2+ t^2\sin^2 k\sin^2\phi})
\end{pmatrix},\nonumber\\
&\ & uh_{2}u^{\dagger}\nonumber\\ &=&-t\cos k\cos\phi \nonumber\\ &\ &+
\begin{pmatrix}
\sqrt{w^2+ t^2\sin^2 k\sin^2\phi} &
0
\\
0&-\sqrt{w^2+ t^2\sin^2 k\sin^2\phi})
\end{pmatrix},\nonumber\\
\end{eqnarray} 
 and the hybridization has the form
 \begin{eqnarray}
&\ &u_{2}Vu_{2} \nonumber\\
&=&\frac{i}{2}(\gamma_{A} f_{y}+\gamma_{B} f_{y})(\cos \theta-i\sin\theta\sigma_{y})\nonumber\\
&\ &+\begin{pmatrix}
- cw & -\frac{i}{2}(\gamma_{A} f_{y}-\gamma_{B} f_{y})\\
-\frac{i}{2}(\gamma_{A} f_{y}-\gamma_{B} f_{y})& cw 
\end{pmatrix}\label{eq:efgap}
%\begin{pmatrix}
%\frac{i}{2}(\gamma_{A} f_{y}+\gamma_{B} f_{y})\cos\theta - cw &
%-\frac{i}{2}( (\gamma_{A} f_{y}-\gamma_{B} f_{y}) + (\gamma_{A} f_{y}+\gamma_{B} f_{y})\sin\theta )
%\\
%-\frac{i}{2}( (\gamma_{A} f_{y}-\gamma_{B} f_{y}) - (\gamma_{A} f_{y}+\gamma_{B} f_{y})\sin\theta )
%&\frac{i}{2}(\gamma_{A} f_{y}+\gamma_{B} f_{y})\cos\theta + cw 
%\end{pmatrix}.
\end{eqnarray} 
The offdiagonal element gives the gap of the effective Hamiltonian:
 \begin{eqnarray}
&\ &[u_{2}Vu_{2}]_{1,2}\nonumber\\
&=& -\frac{it}{2}(\gamma_{A} +\gamma_{B}  )(\cos k\sin\phi+\sin k\cos\phi\sin\theta)\nonumber\\
&\ &- \frac{it}{2}(\gamma_{A} -\gamma_{B} )( \sin k\cos\phi +\cos k\sin\phi\sin\theta)
\end{eqnarray} 
\section{Calculation for the instanton action in the large magnetic field}\label{app:cal}
In the large magnetic field, we have
 \begin{eqnarray}
iS 
&=&\frac{2}{F}\int_{0}^{y_{0}} \mathrm{d }y E_{iy}\nonumber\\
%%%
&=&\frac{2}{F}\sinh y_{0}\cos\phi - \frac{2}{F} \int_{0}^{y_{0}}\mathrm{d }y \sqrt{w^2-t^2\sin^2\phi\sinh^2 y}.\nonumber\\
\end{eqnarray}
Then, by putting $\sqrt{1-\rho^2\sinh^2 y} = \sin\lambda$ ($\lambda\in\left[0,\frac{\pi}{2}\right]$), we have $y =\mathrm{arcsinh}\frac{\cos\lambda}{\rho}$, and the second term is given as
 \begin{eqnarray} 
&\ &\frac{2w}{F} \int_{0}^{y_{0}}\mathrm{d }y \sin\lambda(y)\nonumber\\
&=&
\frac{2w}{F} y_{0} \sin\lambda_{0}-\frac{2w}{F} \int_{0}^{y_{0}}\mathrm{d }y \frac{\mathrm{d }\lambda}{\mathrm{d }y}~y\cos\lambda(y)\nonumber\\
&=&
\frac{2w}{F} y_{0} \sin\lambda_{0}-\frac{2w}{F} \sin\lambda_{0}\mathrm{arcsinh}\frac{\cos\lambda_{0}}{\rho}\nonumber\\
&\ &
+\frac{2w}{F\rho} \int^{\frac{\pi}{2}}_{\lambda_{0}}\mathrm{d }\lambda\frac{\sin^2\lambda}{\sqrt{1+\left(\frac{\cos\lambda}{\rho}\right)^2}},
\end{eqnarray}
with $\sin\lambda_{0}= \sqrt{1-\rho^2\sinh^2 y_{0}} $ and $\rho= \frac{t\sin\phi}{w}$.
The above two first terms cancel each other, and we have
 \begin{eqnarray} 
&\ &\frac{2w}{F} \int_{0}^{y_{0}}\mathrm{d }y \sqrt{1-\rho^2\sinh^2 y} 
\nonumber\\
&=&
\frac{2w\sqrt{\rho^2+1}}{F} \nonumber\\
&\ &\times\int^{\frac{\pi}{2}}_{\lambda_{0}}\mathrm{d }\lambda \left(
\frac{1}
{\sqrt{1-\frac{1}{\rho^2+1} \sin^2\lambda }}
- \sqrt{1-\frac{1}{\rho^2+1} \sin^2\lambda }\right).\nonumber\\
\end{eqnarray}
This is expressed as
 \begin{eqnarray} 
&\ &\frac{2w}{F} \int_{0}^{y_{0}}\mathrm{d }y \sqrt{1-\rho^2\sinh^2 y} 
\nonumber\\
&=&\frac{2w\sqrt{\rho^2+1}}{F} \left[F\left(\frac{\pi}{2}, \sqrt{\frac{1}{\rho^2+1}}\right)-F\left(\lambda_{0}, \sqrt{\frac{1}{\rho^2+1}}\right) \right]
\nonumber\\
&\ & -\frac{2w\sqrt{\rho^2+1}}{F}\left[
E\left(\frac{\pi}{2}, \sqrt{\frac{1}{\rho^2+1}}\right)-E\left(\lambda_{0}, \sqrt{\frac{1}{\rho^2+1}}\right)\right]\nonumber\\
\end{eqnarray}
where
 \begin{eqnarray} 
F\left(a, Q \right)\equiv \int^{a}_{0}\mathrm{d}\lambda\frac{1}{\sqrt{1-Q^2\sin^2\lambda}},\\
E\left(a, Q \right)\equiv \int^{a}_{0}\mathrm{d}\lambda\sqrt{1-Q^2\sin^2\lambda}.
\end{eqnarray}

By using $y_{0}=\mathrm{arcsinh} \sqrt{\left(\frac{w}{t}\right)^2-\cos^2\phi} $, we have $\sin\lambda_{0} =\cos\phi\sqrt{1+\left(\frac{t\sin\phi}{w}\right)^2}$
 \begin{eqnarray}
iS 
&=&\frac{2t}{F}\int_{0}^{y_{0}} \mathrm{d }y E_{iy}\nonumber\\
&=& \frac{2}{F}\int_{0}^{y_{0}} \mathrm{d }y \left(-\cosh y\cos \phi+\sqrt{w^2-t^2\sin^2\phi\sinh^2 y}\right)
\nonumber\\
%%%
&=&\frac{2t\cos\phi}{F}\sqrt{\left(\frac{w}{t}\right)^2-\cos^2\phi} \nonumber\\
&\ &- \frac{2w\sqrt{\rho^2+1}}{F} \nonumber\\
&\ &\times\int_{\lambda_{0}}^{\frac{\pi}{2}}\mathrm{d}\lambda \partial_{\lambda}\left[F\left(\lambda, \sqrt{\frac{1}{\rho^2+1}}\right)+E\left(\lambda, \sqrt{\frac{1}{\rho^2+1}}\right)
 \right].\nonumber\\
\end{eqnarray}

For $iS_{1}$, we have
 \begin{eqnarray}
iS_{1}
&=& -\frac{2w}{F}\int_{0}^{\Tilde{y}_{0}} \mathrm{d }y \sqrt{1-\rho^2\phi\sinh^2 y}
\nonumber\\
%%%
&=&-\frac{2w\sqrt{\rho^2+1}}{F} \left[F\left(\frac{\pi}{2}, \sqrt{\frac{1}{\rho^2+1}}\right)-E\left(\frac{\pi}{2}, \sqrt{\frac{1}{\rho^2+1}}\right)
 \right],\nonumber\\
\end{eqnarray}
with $\Tilde{y}_{0}=\mathrm{arcsinh}\frac{1}{\rho}$.
Then, we have
 \begin{eqnarray}
&\ &iS -iS_{1} \nonumber\\
&=& \frac{2t\cos\phi}{F}\sqrt{\left(\frac{w}{t}\right)^2-\cos^2\phi} 
\nonumber\\ &\ & + \frac{2w\sqrt{\rho^2+1}}{F}\left[F\left(\lambda_{0}, \sqrt{\frac{1}{\rho^2+1}}\right)
-
E\left(\lambda_{0}, \sqrt{\frac{1}{\rho^2+1}}\right)
 \right]\nonumber\\
\end{eqnarray}
\section{Transition probability without virtual scattering at $k=0(\pi)$}\label{ap:transpi}
\begin{figure}[htbp]
 \begin{center}
 \includegraphics[width=80mm]{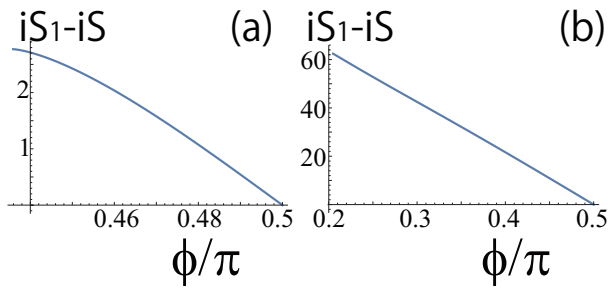}
 \caption{Plot of $iS_{1}-iS$ as a function of $\phi (\geq \phi_{0})$ for $w=0.2$(a) and $w=0.8$(b) at $F=0.01$.
 $iS_{1}-iS\geq 0$ decreases by the increase of $\phi$. $iS_{1}$ is much larger than $iS$ in a wide range of $\phi$,
 and the transition probability by the contribution from $iS_{1}$ is very small.
} 
\label{fig:sdef}
% \label{fig:RTfig}
\end{center}\end{figure}

We consider the transition on $k=0(\pi)$ with the energy difference $E_{k_{0}}^{+}-E_{k_{0}}^{-}=2w$ in the large $\phi$ region.
The action of the transition is given as
 \begin{eqnarray}
&\ &iS_{1}\nonumber\\
&=&-\frac{2w\sqrt{\rho^2+1}}{F} \left[F\left(\frac{\pi}{2}, \sqrt{\frac{1}{\rho^2+1}}\right)-E\left(\frac{\pi}{2}, \sqrt{\frac{1}{\rho^2+1}}\right)
 \right].\nonumber\\
\end{eqnarray} 
 $iS_{1}\leq iS$, and at $\phi=\frac{\pi}{2}$ we have $iS_{1}=iS$.
In Fig.\ref{fig:sdef}, we show the calculation results of $iS_{1}-iS$ as a function of $\phi$ for $w=0.2$(a) and $w=0.8$(b) at $F=0.01$.
Since the transition probability is given by $P_{1}\sim \mathrm{e}^{2iS_{1}}$, when $\phi$ is close to $\frac{\pi}{2}$, the transition should be taken into account. 

%\newpage %Just because of unusual number of tables stacked at end
%\bibliography{apssamp}% Produces the bibliography via BibTeX.

%\bibtem{Goldsmid}
%H. J. Goldsmid ``{\it Thermoelectric Refrigeration}'' (Plenum, New York, 1964).
%\bibitem{Xia}
%Y.Xia {\it et al.}, Nature physics, \texbf{5} 398, (2009).
\end{document}